\documentclass[10pt,journal,compsoc]{IEEEtran}

\usepackage{graphicx}
\usepackage{xcolor}
\usepackage{xspace}
\usepackage{comment}
\usepackage{framed}
\usepackage[strict]{changepage}

\definecolor{formalshade}{rgb}{0.95,0.95,1}

\newenvironment{formal}{%
  \MakeFramed{\advance\hsize-\width\FrameRestore}%
  \noindent\hspace{-4.55pt}
  \begin{adjustwidth}{}{7pt}%
  \vspace{2pt}\vspace{2pt}%
}
{%
  \vspace{2pt}\end{adjustwidth}\endMakeFramed%
}

\newcommand{\MyBox}[1]{\begin{formal}\emph{#1}\end{formal}}

\newcommand{\totalSurvey}{44\xspace}
\newcommand{\totalInvitedInterviews}{eleven\xspace}
\newcommand{\totalRepliedInterviews}{nine\xspace}
\newcommand{\totalInterviews}{seven\xspace}

\begin{document}

\title{How Trans-Inclusive are Hackathons?}

\author{Rafa Prado\textsuperscript{1}, Wendy Galeno\textsuperscript{2}, Kiev Gama\textsuperscript{1}, Gustavo Pinto\textsuperscript{2} \\ UFPE\textsuperscript{1}, Recife, Brazil ~~~~~~ UFPA\textsuperscript{2}, Belém, Brazil

}


\IEEEtitleabstractindextext{%
\begin{abstract}
Hackathons are fun! People go there to learn, meet new colleagues, intensively work on a collaborative project, and mix pizza with energy drinks. However, for transgender community and other minorities, hackathons can have an uncomfortable atmosphere. Some transgender and non-conforming people that, although enjoying hackathons, decided not to participate anymore, afraid of LGBQTPhobia and other discomforts. In this paper we surveyed and interviewed \totalSurvey transgender \totalInterviews hackathon participants. By understanding their needs and challenges, we introduce five recommendations to make hackathons more inclusive.
\end{abstract}

}

\maketitle

%
\IEEEpeerreviewmaketitle

\section{Introduction}

A hackathon is ``a highly engaging, continuous event in which people in small groups produce a working software prototype in a limited amount of time''~\cite{komssi2014hackathons}.
Hackathons became an alternative venue to engage students and practitioners alike in programming activities, while playing a complementary role in the learning and practice of programming skills~\cite{warner2017,nandi2016hackathons}. Hackathons could also foster technical skills that are essential for landing in the tech industry~\cite{pe2019hackathon}.

Unfortunately, the culture around hackathons is still not very welcoming for everyone.  Hackathons ``embody a type of geeky environment that implicitly excludes women and underrepresented minorities''~\cite{warner2017}.
These events can also be particularly harsh for members of the LGBTQIA+ community. Since LGBTQIA+ students are underrepresented in computer science~\cite{stout2016lesbian}, members of this community might have a hard time to find their peers and their space in an event that values communication and collaboration. Due to the low sense of belonging, the absence of other LGBTQIA+ colleagues, and the lack of supportive informal environment, hackathons may actually become a hostile environment~\cite{warner2017}.

Although there are isolated efforts to create more inclusive hackathons~\cite{richard2015stitchfest}, events that explicitly welcome transgender and gender non-conforming (TGNC) people, a subset of the LGBTQIA+ community, represent a rather small fraction~\cite{kumar2019}. This lack of inclusion ultimately impacts on underrepresented groups, which might face fewer opportunities for learning, networking, and jobs~\cite{warner2017}. This effect can be intensified in the transgender community, who has higher unemployment and poverty rates compared to the cisgender population~\cite{winter2016transgender}.

Here we shed some light on the challenges and hurdles that members of the TGNC community might face when participating in hackathons. We conducted a survey with \totalSurvey participants to gauge a broader perception of their challenges and needs. We then conducted \totalInterviews semi-structured interviews to deepen our findings. We then propose five recommendations that could further improve hackathons to better accommodate this so far overlooked community.

\section{Gender: much more than men/women}

\emph{Transgender} (or simply \emph{trans}) refers to ``people who move away from the gender they were assigned at birth, people who cross over (trans-) the boundaries constructed by their culture to define and contain that gender''~\cite{ford2019remote}.
\emph{Gender identity} refers to how one perceives herself, and it is not associated to sexual attraction.
We call \emph{cisgender} (or simply \emph{cis}) when the gender identity aligns with their birth-assigned gender~\cite{transgender-guide}. Note that \emph{straight} is not an opposite of trans; a transgender person can be heterossexual too. When someone defines our society as \textit{cisnormative}, it indicates that our commonsense is to accept only cisgender behavior, and marginalize people that do not follow it~\cite{blockett2017think}.

A transgender person can be also \emph{binary} or \emph{non-binary}. Non-binary trans people are ``individuals whose identity is not exclusively man or woman. While some non-binary individuals identify as both men and women, others have identities that are on the spectrum between man and woman, a different gender entirely, or do not identify with any gender''~\cite{kieran2018definition}. Non-binary can be used as identity, but is more an umbrella term that serves to group identities like \emph{genderfluid} (i.e., the identity of the person flows between other identities) and \emph{agender} (i.e., the absence of a gender).

Gender \emph{non-conformity} is when your \emph{gender expression} (i.e., the way of dress, mannerisms, pronouns, and other characteristics) does not conform to stereotypical gender expectations for their assigned gender~\cite{transgender-guide}. For example, a masculine cisgender woman can define herself as a gender non-conforming person, because she has gender expression associated to masculine gender, but stills identifying herself as a woman.

Another common term is \textit{queer}. ``Queer is a complex word with many different definitions, and in the context of trans communities, it must be recognized as a valid identity term''~\cite{transgender-guide}. This term can be used to define sexual attraction, gender identity, and gender expression. One needs to understand more about the context when, for example, queer is cited in an interview ~\cite{transgender-guide}.

\section{Method}



\subsection{Survey}

We set up the survey as an on-line questionnaire
with 22 questions, nine of which were open-ended. The survey had four sections: (1) gender identity and education level\footnote{The gender identity question was open-ended, so participants could identify themselves as they wanted. We also asked if the participants identify themselves as a trans or a non-conforming person.}, (2) experience in participating in hackathons, (3) perception about TGNC people on the inclusion of these events, (4) any personal comment to add. To make participants more comfortable in answering the survey, all questions were optional. The estimated time to complete the survey was 10 minutes.
We had different questions for TGNC and cis participants. For TGNC participants, we asked about 
previous uncomfortable situations they had experienced, their concerns about participating, and what would make them more comfortable to participate in hackathons. For cisgender participants, we asked their perceptions about transgender people participating in hackathons, and their opinion about a more gender-inclusive hackathon event.

We posted a call for participating in specific mediums, such as hackathons mailing lists, on-line inclusive tech forums, Twitter, and in LGBTQI+ Students Alliances groups. We also sent this questionnaire directly to transgender colleagues who we know that participate in hackathons, and asked them to forward it to other transgender colleagues they might have. We received \totalSurvey responses.

We qualitatively analyzed the open-ended answers~\cite{SeamanTSE1999}. We took care to group the answers from cis, transgender, or non-conforming participants. 
On average, our respondents were approximately 24 years old (min: 15, max: 42).
Five participants identified themselves with a non-binary identity (two of them self-identified as genderfluid). From the remaining, 25 are men (2 non-conforming, 4 trans, and 19 cis) and 13 are cis women. One participant identified herself as ``travesti", a latin-american identity that is linked to womanhood, but has its own meaning and should not be translated \cite{pierce2020monster}.
The survey participants are identified as \emph{S}1--\emph{S}\totalSurvey .

\subsection{Interviews}

We invited \totalInvitedInterviews TGNC people to participate in our \emph{semi-structured}  interviews. Among these invitations, \totalRepliedInterviews participants were happy to talk to us, but only \totalInterviews ended up participating.


A non-binary researcher conducted the interviews from July 2020 to November 2020. We followed ethical recommendations for research with transgender participants, such as understanding the transgender context and the usage of nuanced language~\cite{vincent2018studying}.
The interview had four parts: (1) the participant's development experience and their gender identity; (2) their good or bad experiences in hackathons; (3) their opinion and experience with on-line hackathons; and (4) their ideas concerning more inclusive hackathons.
We conducted online (video and audio) interviews in Portuguese. On average, the interviews lasted 39 minutes (min: 17, max: 72). We asked the participant permission to record the audio, which we later transcribed.

\begin{figure*}[t]
    \centering
    \includegraphics[width=\linewidth,
    clip=true, trim= 0px 0px 0px 0px]{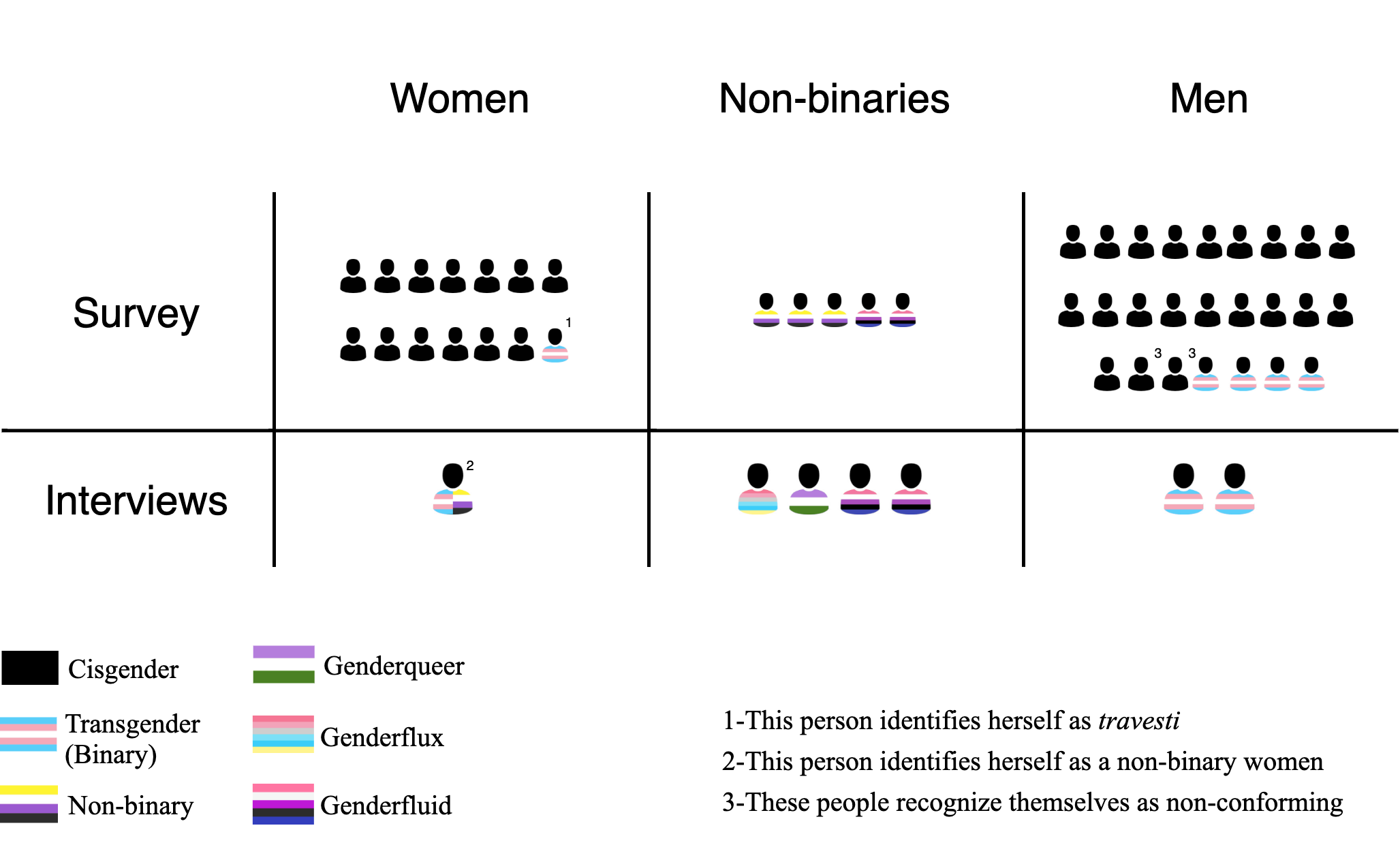}
    \caption{Overview of our participants' gender identities.}
    \label{fig:participants}
\end{figure*}

Similar to the survey, we coded the transcripts following general open coding guidelines~\cite{SeamanTSE1999}. Each transcript, along with the associated recording, was analyzed by two authors. The coding procedure was done independently, followed by conflict resolution meetings. A third researcher revised the process and helped to reach consensus.
The coding procedure was done in Portuguese, whereas the quotes were translated to English.
On average, our respondents were 23 years old (min: 20, max: 27).
One participant is a non-binary trans woman, one is as genderfluid, one is a queer person, two are non-binary, while the remaining two are trans men.
The interview participants are identified as \emph{I}1--\emph{I}7. Figure~\ref{fig:participants} shows an overview of our participants.

\subsection{Limitations}

A first limitation is related to the number of participants in our study.
However, our population is 1) not straightforwardly identified, 2) not easily approached, and 3) not frequently studied. We concur that the population we assembled is small but yet reasonable to draw a first impression about this so far uncharted landscape.
We also focused on Brazilian participants due to our proximity with them; we shortly noticed that many trans developers were not comfortable in participating in interviews, requiring us to first build a trustable relationship through referrals before inviting them to participate in the study.
As a way to understand how trans people are perceived by cis people, we also solicited answers from cis people in the questionnaire (we though did not greatly spread this questionnaire to cis groups).

\section{Findings}

\subsection{Learning and fun}

Our participants mentioned several benefits in participating in hackathons, including 1) acquiring new technical skills, 2) breaking the monotony, 3) meeting new colleagues, or 4) immersing in a short-term group project. Although these benefits have been extensively reported in the literature~\cite{warner2017}, we also noted that hackathons could play a prominent role for transgender developers. For instance, one interviewee mentioned that:

\MyBox{{\bf``}There are several skilled people that need to have visibility, because we [transgender people] don't have many job opportunities{\bf''} (I03, non-binary trans woman)}

However, to take proper advantage of the hackathon, six interviewees mentioned training events as a way to approximate trans people to hackathons.
For instance, S46 explained that hackathons focused on training and providing a friendly collective environment rather than a competitive one is a way to attract more TGNC people
to the event.


\subsection{Hackathons are not planned to be transgender inclusive}

Some of our
participants mentioned bad experiences in participating in hackathons. Five survey respondents reported facing discrimination because of their gender expression, and other two reported facing other kinds of prejudice. Three respondents mentioned facing discrimination (either due to their identity or expression) from the organization staff. Four respondents do not want to participate anymore because of being afraid to suffer from LGBTQphobia (three respondents are also afraid, but do not intend to stop participating). One interviewee taught us that such concerns could co-exist in different ways:

\MyBox{{\bf``}It starts happening on the registration, where I look around at and see 
the specific audience [cisgender men]. It made me feel a certain insecurity. The premises count a lot. For instance, if the event has a gender-neutral bathroom, if the place looks safe, etc.{\bf''} (I02, a trans man)}

These concerns are greatly exacerbated by the sense of loneliness, since transgender people are often the only non-cisnormative person participating. None of our interviewees noticed any other transgender in the hackathons they participated. Only three cisgender survey participants have seen transgender people participating in  hackathons.
Exposing their identity is also a concern for our participants.
Four survey respondents mentioned that they were not able to use a different name when registering for the hackathon. One interview participant (I01, a trans man) also mentioned that: ``\emph{there were hackathons that you had to present your identity card; at the time I had not yet made the gender transition}''

One interviewee (I06) declared that transphobia can happen in different, more subtle ways. The participant declared being as a non-binary person to the event organizers and participants. However, although their group talked respectfully, they did not request them in any challenging way, as fully-capable member, showing an example of hidden discrimination. I06 complemented saying ``\emph{this experience was very traumatic; I may not use it [the gender identity] in the future to avoid this type of situation}''


Finally, five other survey respondents also mentioned that the majority of these events do not have clear policies or code of conduct against LGBTQphobia and gender discrimination. Six out of nine trans survey respondents have seen a code of conduct against LGBTQphobia in hackathons. However, when needed, they reported that it was not applied. Two interviewees (I4, queer and I5, genderfluid) said the organization staff just \emph{verbally} expressed that the event would not tolerate any expression of racism, sexism, or LGBTQphobia.

\MyBox{{\bf``}No! None of the hackathons I participated in had this explicitly, for example: verbal or physical aggression will not be allowed here{\bf''} (I01, a trans man)}

\subsection{Online hackathons to the rescue?}

Since remote work allows identity disclosure, providing a sort of safe space to transgender developers~\cite{ford2019remote}, we believed a similar comfort zone could be created in the case of online hackathons. Surprisingly, only two respondents mentioned that going online would mitigate the problems reported by the survey respondents (eight participants answered `maybe' and two participants answered `no').
I07 punctuated that ``\emph{If most people don't know how to deal with it personally, it would not change in a virtual scenario. For instance, a visibly non-conformist person was added in group [in an online event]. During the presentation one of the members said ``She, he, I don’t know"}''.


Other interviewees also shared some concerns with online hackathons. For instance, I02 (a trans man) mentioned that \emph{``there are many people who do not feel safe even at home. Going to an event where you will be with people for 48--72 hours, and have their support, this atmosphere becomes much safer than their own home"}. Additionally, I02 (a trans man) also pointed out that online hackathons would miss in-person interactions, which is essential for such events.

\subsection{Hackathons for everyone!}

When we asked about how hackathons could be more inclusive, eight survey respondents stated the need to introduce clear incentive and inclusion policies for transgender people. Moreover, S03 (cis woman) highlighted that it is equally important to make sure these policies are being strictly followed and  provide means to denounce participants that are not obeying such policies. Allowing participants to chose their own pronouns
was also a recurring comment by our survey respondents. Also, all interview participants and five survey respondents mentioned that having LGBTQI+ people as mentors or staff would be helpful to make hackathons more inclusive. 
Campaigns promoting transgender participation in hackathons are also important:

\MyBox{{\bf``}Campaigns make us feel that the event values our participation. It also helps people who have a more closed mind to understand that hackathons are not a space to propagate hate, as some of them were used to do.{\bf''}(S17, a trans man)}

Inclusive hackathons might also mitigate other problems for the TGNC community: lack of professional opportunities. The networking
could help trans people to find prospective jobs. I07, a non-binary person, said \emph{``The tech industry is on constant growth but LGBTQIA+ people are still on the margin. We are just as competent as straight white cis men. When hackathons showcase members of the TGNC community, they give opportunity to them. Everything is an opportunity.''}

\section{Five recommendations}

We now propose five evidence-based recommendations for a gender-inclusive hackathon event.
The ultimate goal is to create a safe and welcoming space for TGNC community, while allowing more people to participate (diversity) and sharing a good experience (inclusion).

\subsection*{REC\#1: Start with a gender-inclusive organizing team}
Although
TGNC people are participating as organizers/mentors, their participation is still scarce.
Trans people in the organization would not only assist creating additional inclusion measures, but could also help building a safer space for other trans participants. Also, by knowing that there are TGNC people as staff (especially mentors, who establish close contact with the participants to give tech guidance), trans participants might be even more comfortable in participating.

\subsection*{REC\#2: Foster inclusive communication}
The use of an inclusive language (previous, during, and after the event) is essential to welcome all audiences, including TGNC.
In the registration form, allow participants to fill their preferred names (avoid requesting their registration name) and pronouns. Rather than having a pre-fixed set of identities, leave open the gender-identity field. Provide badge cards in which they can put their names and pronouns. For online events, make pronouns visible on the participants' profile. For organizers: make sure that participants, mentors, and staff are respecting the information trans people provided.

\subsection*{REC\#3: Make safety visible}

A hackathon must have a code of conduct, which clearly states that LGBTQphobia and other discrimination acts (not only from participants) are not tolerated. The staff must also verbally reinforce the code of conduct during the event, which would remind that everyone (including mentors and staff) needs to be respectful with each other. Still, it is important to create a safe way to denounce bad behavior, 
so any code of conduct violation can result in expulsion from the event.

\subsection*{REC\#4: Provide good working conditions to participants}
Some trans people might not feel capable of attending a hackathons due to the lack of technical knowledge or experience. Organizers could then create pre-hackathon events focused on inclusion and training, 
giving them a chance to feel more comfortable to deal with proposed challenges. The financial situation of the trans community, which is still a marginalized group, is also an issue.
Organizers could offer proper equipments or propose projects that could be done without using high performance technologies. 

\subsection*{REC\#5: Showcase trans people in the event}
Hackathons
can be platforms for people to achieve professional opportunities. When tech companies are invited to visit a hackathon
, the participants' talents can be recognized and can be invited to selection processes, which would hardly have happened under normal conditions. 
By highlighting trans talents, hackathons can help empowering the whole TGNC community, supporting their fight for equity.

\bibliographystyle{IEEEtran}
\bibliography{bibliography.bib}





\end{document}